\documentstyle[12pt]{article}

\newcommand{\be}{\begin{equation}}
\newcommand{\ee}{\end{equation}}
\newcommand{\mpl}{ {M_{\rm pl}}} 
\newcommand{\mpln}{ {M_{{\rm pl}\ast}}} 
\newcommand{\lpln}{ {\ell_{{\rm pl}\ast}}} 
\newcommand{\mnew}{ {M_{\rm qg}}} 
\newcommand\pp{\parshape 2 0.0truecm 14.25truecm 2truecm 12.25truecm}

\begin{document}

\baselineskip 20pt 

\centerline{\bf PROTON DECAY, BLACK HOLES, and LARGE EXTRA DIMENSIONS}

\bigskip 
\centerline{\bf Fred C. Adams, Gordon L. Kane, and Manasse Mbonye} 
\bigskip 
\centerline{Michigan Center for Theoretical Physics} 
\centerline{Physics Dept., University of Michigan, Ann Arbor, MI 48109, USA}
\bigskip 
\centerline{\bf Malcolm J. Perry} 
\bigskip 
\centerline{DAMTP, University of Cambridge, Wilberforce Road, Cambridge, 
CB3 0WA, England} 
\bigskip 
\centerline{\it 29 December 2000} 
\bigskip 

\begin{abstract} 
We consider proton decay in theories that contain large extra
dimensions. If virtual black hole states are allowed by the theory, as
is generally the case, then proton decay can proceed via virtual black
holes. The experimental limits on the proton lifetime place strong
constraints on the quantum gravity scale $\mnew$ (the effective Planck
mass). For most theories, this implies a lower bound of $\mnew >
10^{16}$ GeV. The corresponding bound on the size of large extra
dimensions is $\ell < 10^{6/n} \times 10^{-30}$ cm, where $n$ is the
number of such dimensions. Regrettably, for most theories this limit
rules out the possibility of observing large extra dimensions at
accelerators or in millimeter scale gravity experiments. Conversely,
proton decay could be dominated by virtual black holes, providing an
experimental probe to study stringy quantum gravity physics.
\end{abstract}

\bigskip 

\bigskip 
\newpage 
\centerline{\bf I. INTRODUCTION} 
\bigskip  

In conventional versions of string theory (M theory), the string
energy scale, the Planck mass, and the unification scale are roughly
comparable and are relatively close to the standard value of the
Planck mass, $\mpl$ $\sim 10^{19}$ GeV. Recently, however, the
possibility of a much smaller string scale and large extra dimensions
-- perhaps large enough to be observable in particle accelerator and
gravity experiments -- has sparked a great deal of interest [1--11].
As is well known, one of the many constraints on such theories with
large extra dimensions is the rate of proton decay. The current
experimental limit on the proton lifetime depends on the particular
decay mode under study [12], but for the most interesting channels the
bound is approximately $\tau_P > 10^{33}$ yr [13]. In the context of
theories with large extra dimensions, the existing theoretical
literature includes various papers which present mechanisms to
suppress ``ordinary'' GUT scale proton decay, i.e., decay driven by
intermediate bosons that mediate baryon number non-conservation.  In
this case, one can invent extra symmetries to get rid of those
unwanted processes and thereby suppress the proton decay rate below
the experimental limit [4--7]. In this present paper (see also Refs.
[4,11]), we address the idea of virtual black holes acting as the
intermediate particles as they do in gravitationally induced proton
decay [14--18]. In particular, we find lower bounds on the quantum 
gravity scale $\mnew$, upper bounds on the size $\ell$ of the extra 
dimensions, and limits from higher order proton decay processes with 
$\Delta B > 1$. 

This topic has some urgency: If the relevant scale for quantum gravity
is as low as 1 TeV, for example, then quantum gravity effects could in
principle be observed in existing (or upcoming) accelerators, or as
deviations from Newtonian gravity.  One of the proposed experimental
signatures of quantum gravity would be virtual black holes
[8,19]. Unfortunately, the same virtual black holes that could be
observed could also drive proton decay at a rapid rate, larger than
allowed by existing experimental bounds on the proton lifetime. As a
result, gravitationally induced proton decay must be highly suppressed
in any theory of quantum gravity that accommodates low energy scales
and thereby allows large extra dimensions; possible mechanisms have
been suggested to provide such suppression and remain under study
[4,5,6,11]. However, in a viable theory such a suppression must allow
a way to generate a baryon asymmetry and neutrino masses; these can be
linked by $B-L$ conservation in many theories. So far, that has not
been achieved to our knowledge. The burden of proof must be, to a
large extent, placed on any comprehensive new approach, so that it
demonstrates that it is not inconsistent with these constraints. The
arguments of this paper thus suggest interesting constraints on
approaches to quantum gravity that allow for virtual black hole states
at low energy scales, such as recent approaches invoking millimeter or
TeV size dimensions.

For another new approach, the Randall-Sundrum case [20,21], the
Standard Model (SM) masses appear at about the TeV energy scale while
the original mass parameter of the theory can remain near the (old)
Planck scale $\mpl$.  The energy scale for quantum gravity effects and
virtual black holes may remain large ($\sim \mpl$) and hence this
class of theories might evade the bounds of this paper. However, in
most versions of Randall-Sundrum ideas where Kaluza-Klein (KK) states
are at the TeV scale, and where black hole sizes are determined by the
KK masses, our arguments should apply.

\newpage 
\centerline{\bf II. PROTON DECAY IN 4 DIMENSIONS} 
\bigskip 

Let's first review the picture of gravitationally induced proton decay
using three spatial dimensions and the traditional value of the Planck
mass ($\mpl \sim 10^{19}$ GeV). In any quantum theory, one expects to
find vacuum fluctuations associated with the fundamental excitations
of the theory. In electrodynamics, for example, electron-positron
pairs can form directly out of the vacuum (for a short time). Such
processes can be observed indirectly by many quantum phenomena (e.g.,
in the Casmir effect). In general, however, one must include in the
vacuum processes all possible excitations of the theory, e.g., the
production of proton-antiproton pairs, or even monopole-antimonopole
pairs. These processes are generally highly suppressed relative to the
electron-positron amplitudes by virtue of their correspondingly large
masses. In gravitation, one therefore expects not only to find virtual
gravitons playing a role, but also virtual black holes. Although
present uncertainties in quantum gravity theory prevent reliable
calculations, even in the context of string theory, we can use a
semi-classical calculation as a starting point to study such
phenomena.

The standard arguments for virtual black holes lead to the idea that
spacetime must be filled with tiny Planck mass black holes with a
density of roughly one per Planck volume [22,23]. These microscopic
virtual black holes then live roughly for one Planck time. This
picture of the spacetime vacuum is often called the {\it spacetime
foam} (a generalization of this argument for higher dimensions is
sketched in the Appendix). Since black holes and the Standard Model
itself are presumed not to conserve baryon number, these virtual black
holes contribute to the rate of proton decay through their
gravitational interaction. In this setting, a proton is considered to
be a hollow sphere of radius $R \sim m_P^{-1} \sim 10^{-13}$ cm that
contains three (valence) quarks. Suppose that two of these quarks fall
into the same black hole at the same time (the quarks must be
pointlike compared to the black hole scale for this argument to hold).
Since the black hole will evaporate predominantly into the lightest
particles consistent with conservation of charge, energy, and angular
momentum, this process effectively converts the quarks into other
particles; only rarely will the same quarks come out that originally
went into the black hole. The output particles will often be
antiquarks and leptons, and hence baryon number conservation is
generally violated. In other words, quantum gravity introduces an
effective interaction leading to many final states, including
processes of the form 
\be 
q + q \to {\bar q} + \ell \, , 
\label{eq:interaction} 
\ee 
where the final state can include any number additional particles
(gravitons, gluons, photons, neutrinos, etc.) and where the resulting
antiquark will generally hadronize (e.g., to $\pi^0$).  These
interactions can be regarded as four-Fermi interactions whose coupling
strength is determined by the Planck mass. Such processes are mediated
by black holes and can violate conservation of baryon number.  Notice,
however, that these processes cannot be mediated by gravitons alone
because such interactions conserve both electric charge and baryon
number. Notice also that we are implicitly adopting the Hawking 
picture of black hole evaporation (which is information non-preserving). 

The probability of two quarks being within one Planck length
($\ell_{\rm pl} \sim 10^{-33}$ cm) of each other inside a proton is
about $(m_P/\mpl)^3 \sim 10^{-57}$. This value represents the
probability per proton crossing time $\tau \sim m_P^{-1} \sim
10^{-31}$ yr, if we assume that the particles move at the speed of
light.  In order for an interaction to take place (such as equation 
[\ref{eq:interaction}]), a virtual black holes must be present at the
same time that the two quarks are sufficiently near each other.
Including this effect reduces the overall interaction probability by
an additional factor of $m_P/\mpl$.  Converting these results into a
time scale for proton decay [14--18], we find an estimated proton 
lifetime of
\be
\tau_P \sim m_P^{-1} \Bigl( {\mpl \over m_P} \Bigr)^4 \, \sim 
10^{45} \, {\rm yr} \, .  
\label{eq:tgrav} 
\ee
For comparison, if the proton is unstable through some process 
operating at the (nonsupersymmetric) unification (GUT) scale 
$M_X$ [12], the corresponding time scale for proton decay becomes 
\be 
\tau_P \sim 10^{30} {\rm yr} 
\Bigl( {M_X \over 10^{15} {\rm GeV}} \Bigr)^4 \, . 
\label{eq:gut}
\ee
Thus, the proton lifetime expected from virtual black hole processes
(equation [\ref{eq:tgrav}]) is the same as that for GUT scale
processes (equation [\ref{eq:gut}]) in the limit that the unification
scale $M_X$ approaches the Planck scale $\mpl$ and the coupling
becomes of order unity. For completeness, we note that in
supersymmetric theories the unification scale can be somewhat higher
than in grand unified theories without SUSY; in this case, the proton
lifetime can be as long as $\tau_P$ = $10^{33}-10^{34}$ yr, consistent
with current experimental limits.

Most versions of string theory contain black hole states with masses
comparable to the Planck mass (the quantum gravity or string scale).
These black holes play the role of the $X$-boson in proton decay,
independent of any specific argument about virtual black holes or
spacetime foam. Furthermore, we have no reason not to believe the
Hawking formula for the entropy of stringy black holes or their
temperature; as a result, the density of states formula implicitly
used here should be correct in string theory. One remaining
controversy is whether or not the black holes genuinely lose
information. Most particle physicists say they do not, whereas most
relativists say they do. At present, we simply do not know. Notice
that if only $B-L$ is conserved in string theory, then protons will
decay with the rate derived here.

One still might be concerned about suppression of the proton decay
rate due to violation of global conservation laws or information
loss. Because of the t'Hooft anomaly, however, baryon number is not
conserved even in the SM, although electric charge and possibly $B-L$
are. The decay described by equation (\ref{eq:interaction}), e.g.,
conserves these latter quantum numbers. Channels such as this decay,
that could conserve global quantum numbers and perhaps even circumvent
information loss issues, can dominate [10]. As a result, there is no
compelling reason to expect large suppressions of the decay channel.

In equation (\ref{eq:interaction}), the virtual black holes mediating
the interaction appear to act like local quantum objects and thus one
might be concerned that the interaction could be gauged away, much
like what is done to remove unwanted interactions involving
$X$-bosons. Unlike local quantum particles, however, the virtual black
holes in quantum gravity processes are solitonic objects -- they are
less likely to be gauged away because they are extended. In order to
suppress proton decay mediated by virtual black holes, one would need
to make [5,6] baryon number (or an equivalent matter parity) into a
local charge via an exact gauged discrete symmetry that is fully
respected by the true vacuum of the theory. Even then, proton decay is
often only suppressed up to some (possibly quite large) order in the
effective operators. Such a suppression requires very special
arrangements of chiral fermions or other aspects of the theory. For
example, as pointed out by Kakushadze [6], a ${\bf Z}_3$ ``Generalized
Baryon Parity'' cannot accommodate right-handed neutrinos without
adding additional matter because of anomaly cancellation
constraints. It is not an accident that the neutrino sector affects
the efforts to enforce such suppressions; $B$ and $L$ conservation are
related in many models by conservation of $B-L$. To allow majorana
neutrino masses, $L$ cannot be conserved.  As we argue below, however,
because theories with large extra dimensions tend to allow rapid
proton decay via virtual black holes, any viable such theory must
include a strong suppression mechanism.

Some mechanisms have been suggested [11,27] to avoid the proton decay
problem in the presence of large extra dimensions. They typically have
in common the need for a new scale, one that is not given by scale of
quantum gravity and is not determined by conventional (known) scales
such as the electroweak or supersymmetry breaking scales. In
Ref. [11], for example, quarks and leptons are embedded in domain
walls separated by a new scale of order 50 times the distance (or
more) one would expect if the wall parameters were determined by the
quantum gravity scale; if the domain wall thickness was specified only
by the quantum gravity scale, then virtual black hole states would be
large enough to bridge the gap and drive proton decay. If the physical
origin of such a new scale can be identified, and a reason why a
stable separation of quarks and leptons should occur, that would
represent important progress. One additional general constraint on
such ideas is the need to generate a baryon asymmetry in the early
universe (this issue is not addressed in Ref. [11]). Currently, quarks
and leptons are separated by imposing the counter-intuitive
requirement that their Yukawa couplings are of opposite sign, which is
possible but, as yet, unmotivated.

Although it is not yet absolutely proven that baryon number cannot 
be effectively conserved in the presence of black holes, we find it 
unlikely for two reasons: {\bf [A]} Astrophysical black holes seem to
manifestly violate such a conservation law. Imagine compressing a star
containing $N_B \sim 10^{57}$ baryons into a black hole and watching
it radiate away. Because the Hawking temperature is low for most of
its evaporation time, the black hole radiates primarily into photons,
gravitons, and neutrinos; the temperature becomes hot enough to
radiate quarks, protons, or other baryonic particles only after the
mass shrinks by 20 orders of magnitude. Thus, for baryon number to be
conserved, the theory would have to contain extremely unusual objects
with small mass and huge baryon number -- the baryon number to mass
ratio would be 20 orders of magnitude larger than that of the proton.
This case may not be explicitly excluded but it is nonetheless
extremely unlikely; it requires the introduction of a new factor of
$10^{20}$, which is ten million times larger than the dimensionless
number ($M_{GUT}/M_{Weak} \sim 10^{13}$) represented by the hierarchy
problem. (Notice that the black hole may not actually disappear, just
as an electrically charged black hole may not disappear -- it remains
in a BPS-like configuration; the implications of this possibility
remain unclear). {\bf [B]} The observed baryon asymmetry in the
universe argues strongly against absolute conservation of baryon
number. Unless one posits special initial conditions at the Big Bang,
the cosmos had to generate a baryon excess through {\it some} process
that violates conservation of baryon number.

In string theory, a large number of BPS states are known to be extreme
black holes [24], and presumably many other massive string states are
black holes as well. All of these states can mediate decays as in
equation (\ref{eq:interaction}). Because so few known experimental
tests can probe the string nature of quantum gravity theories, we
should turn our argument around: Since string theory is likely to
allow proton decay via virtual black holes and since the string scale
may be as low as the GUT scale, proton decay modes (such as those
explored in this paper) may be a very powerful diagnostic of string
theories. Furthermore, the gravitationally induced channels of proton
decay should be recognizable in experiments from their observed
branching ratios. Thus, proton decay could be a valuable way to study
strong gravitational interactions.  This issue should be studied in
greater detail in future work.

\bigskip 
\bigskip 
\centerline{\bf III. PROTON DECAY WITH LARGE EXTRA DIMENSIONS} 
\bigskip 

We now consider the process of gravitationally induced proton decay in
theories with large extra dimensions. For proton decay driven by
non-gravitational means -- for intermediate particles other than
virtual black holes -- it is possible to enforce symmetries on the
theory to prevent proton decay at overly fast rates [4--7]. In the
case of gravity, however, such suppressions are more difficult,
particularly when quantum fluctuations are large. Working in worlds
with large extra dimensions, Emparan et al. [10] have argued that
black hole evaporation occurs mostly on the brane; this result thus
strengthens our approach since final states with particles lighter
than a proton are not suppressed.

In a theory with large extra dimensions, two effects modify 
the picture of proton decay outlined above:  

{\bf [A]} The most important modification is that the Planck mass
changes. Specifically, the energy scale of virtual black hole
processes changes from $\mpl \approx 10^{19}$ GeV to a lower value
which we denote here as $\mnew$. Because the new quantum gravity scale
$\mnew$ is generally lower than both $\mpl$ and the GUT scale, this
effect acts to reduce the proton lifetime.  In particular, the
Schwarzschild radius for the virtual black holes is given by $R_S$
$\sim$ $\mnew^{-1}$ [8], which determines the cross section and is
much larger than before.

{\bf [B]} If the number of extra large dimensions is $n>0$, then the
geometry of both the proton and the virtual black holes change. In
this context, we let $d \le n$ denote the number of extra dimensions
{\it that the quarks can propagate through}. In most theories with
large extra dimensions, quarks and other SM particles are confined to
the usual 4-dimensional world and cannot freely propagate in the extra
dimensions; for most cases, we thus have $d=0$. In the general case
with $d>0$, the quarks that make up the proton have more dimensions in
which to propagate. With more dimensions, the quarks would be less
likely to encounter each other and hence this effect increases the
proton lifetime. On the other hand, the black holes must be $(4+n)$
dimensional objects and will necessarily live in the additional
dimensions. The black hole interaction cross sections remain of order
$R_S^2 \sim \mnew^{-2}$, however, even when interacting with SM
particles confined to the usual 4-dimensional spacetime (where we 
assume strong coupling). 

Including the above two modifications in estimating the proton decay
rate through virtual black hole processes, we find the proton lifetime
\be 
\tau_P \sim m_P^{-1} \Bigl( {\mnew \over m_P} \Bigr)^{4+d} \, .
\label{eq:taunew} 
\ee 
The current experimental bound on the proton
lifetime [13] can be written in the form 
\be 
\tau_P > 10^{33} \, {\rm yr} \, \equiv m_P^{-1} 
\Bigl( {\Lambda \over m_P} \Bigr)^{4} \, , 
\label{eq:observe} 
\ee 
where we have defined an energy scale $\Lambda \equiv$ 
$(m_P^5 10^{33} {\rm yr})^{1/4}$ $\approx 1.4 \times 10^{16}$ GeV.
Combining the general expression (\ref{eq:taunew}) with the
experimental bound (\ref{eq:observe}), we thus obtain a bound 
on the scale $\mnew$ of quantum gravity:
\be 
\mnew > \bigl( m_P^d \Lambda^4 \bigr)^{1/(4+d)} \, = \, 
10^{64/(4+d)} \, {\rm GeV} \, , 
\label{eq:qgbound} 
\ee 
where we have used $m_P \sim 1$ GeV and $\Lambda \sim 10^{16}$ GeV 
to evaluate the bound in the second equality. This result (when 
applicable) constrains the possibility of having a low quantum gravity
scale that could be observed in present-day or future accelerators. 
For the most likely case $d=0$, where quarks are confined to our
4-dimensional brane, the quantum gravity scale must be comparable to
the (usual) GUT scale, i.e., $\mnew$ $\ge$ $10^{16}$ GeV. For $d=1-2$,
the quantum gravity scale remains quite high. The weakest constraint
arises if $d=7$, which corresponds to the (unlikely) case in which all
of the possible extra dimensions are large and the valence quarks
within the proton are allowed to propagate freely through all
dimensions; in this case, the limit on the quantum gravity scale is
$\mnew >$ 700 TeV. This scale remains interesting in terms of
modifying the hierarchy problems associated with a high quantum
gravity scale ($\mpl \sim 10^{19}$ GeV), but unfortunately it remains
safely out of experimental reach.

We can also find corresponding bounds on the size scales of the 
extra dimensions. This size scale $\ell$ is determined by 
\be
\ell^n = \mpl^2 \mnew^{-(2+n)} \, , 
\ee
where $n$ is the number of extra large dimensions (see Ref. [1--11]). 
For the most likely case with $d=0$, equation (\ref{eq:qgbound}) 
implies the bound 
\be
\ell < (\mpl/\Lambda)^{2/n} \Lambda^{-1} \, . 
\ee
As a result, the ``large'' extra dimensions in such a theory would
actually be rather small, $\ell < 10^{6/n} \times 10^{-30}$ cm.  
These size scales would be impossible to observe in modified 
gravity experiments. 

Other rare decays mediated by virtual black holes, such as $\mu \to e
\gamma$ or neutrino disappearance, may also provide limits on large
extra dimensions and low quantum gravity scales.  We leave the study
of these effects to a future analysis.

Thus far, we have only considered processes which lead to the decay of
a single proton, i.e., processes with $\Delta B$ = 1. However, many of
the possible suppression mechanisms for proton decay forbid $\Delta B$ 
= 1 decays, but allow larger $\Delta B$. For example, $\Delta B$ = 2
can appear in neutron-antineutron transitions.  We can immediately
generalize the expression for the proton lifetime (eqs. [\ref{eq:tgrav} 
-- \ref{eq:gut}]) for the case of $\Delta B = N$ (see Ref. [28]), 
\be
\tau_P \sim m_P^{-1} \alpha^{-2N} 
\Bigl( {\mnew \over m_P } \Bigr)^{4N} \, \sim \, 10^{33} {\rm yr} 
\, \, 10^{64(N-1)} \, \Bigl( {\mnew \over 10^{16} {\rm GeV}} 
\Bigr)^{4N} \, , 
\label{eq:higher}
\ee
where $\alpha$ is the coupling constant and where we have taken the
likely case of $d=0$. For example, if we use the scale $\mnew$ = 1 TeV
for quantum gravity, the proton lifetime for $N$ = 2 is only a few
seconds; for $N$ = 3, the proton lifetime is only $\tau_P \sim 10^5$
yr. We note that processes with $\Delta B \ge 2$ require the protons
to be near each other. Such processes will thus take place in large
nuclei (e.g., iron) and in compact stellar objects (e.g., neutron
stars) but free protons in interstellar space would not be affected.

We also obtain the corresponding bound on the quantum gravity 
scale $\mnew$, 
\be
\mnew > m_P (\Lambda/m_P)^{1/N} \approx 10^{16/N} \, {\rm GeV} \, . 
\label{eq:delb} 
\ee 
For the case of $N$=2, for example, our bound becomes $\mnew > 10^8$
GeV; for $N$=3, the bound becomes $\mnew > 2 \times 10^5$ GeV. Thus, 
higher order proton decay processes (with $\Delta B \ge 2$) also 
place interesting limits on the quantum gravity scale $\mnew$. 

\bigskip 
\bigskip 
\centerline{\bf IV. SUMMARY} 
\bigskip  

This paper argues that gravitationally induced proton decay -- virtual
black hole processes that violate baryon number conservation -- should
be taken seriously as they imply strong constraints on theories of
quantum gravity with large extra dimensions. In particular, this
analysis suggests that the observed absence of proton decay via
virtual black holes puts a lower limit on the quantum gravity scale
$\mnew$ and a corresponding upper limit on the size $\ell$ of large
extra dimensions. In the weakest (and unlikely) case in which quarks
propagate in $n=d=7$ large extra dimensions, the limit is $\mnew$ $>$
700 GeV. This bound rapidly increases to $\mnew > 10^{16}$ GeV for any
number $n$ of large extra dimensions, if quarks move only in 3 spatial
dimensions ($d=0$) as is generally required for theories to retain the
usual SM physics [1--11,25]. The corresponding bound on the size scale
of the extra dimensions is $\ell < 10^{6/n} \times 10^{-30}$ cm. The
bounds for $\Delta B > 1$ processes are weaker, but the quantum 
gravity scale is still highly constrained (eq. [\ref{eq:delb}]).

Because the required interactions with black holes are very general,
this limit is robust and will not be affected by the domination of
specific decay channels. It could be modified if quark sizes (perhaps
set by the string scale) are larger than the Planck size, but this
does not occur in most approaches. Our limit may not apply if the
generally accepted picture of spacetime foam -- every Planck volume of
spacetime typically contains a virtual black hole for a Planck time
(see the Appendix) -- is not valid, or if virtual black hole states
are charged under some conserved (including quantum corrections)
discrete gauge symmetry (as discussed earlier). Other possible
mechanisms to suppress virtual black hole effects have been suggested
[11] and should be studied further. Any acceptable mechanism should
allow an explanation of the cosmic baryon asymmetry and majorana
neutrino masses. We feel that this challenge remains a serious issue.

In general, if this limit turns out to be applicable, the quantum
gravity scale must be high enough to remove much of the original
motivation for large extra dimensions. Unfortunately, this argument
would rule out observable effects at colliders and in millimeter-scale
gravity experiments; this issue is thus of vital importance and must
be dealt with more fully than it has been so far. Even with this new
constraint, however, the quantum gravity scale $\mnew$ could still be
somewhat lower than before ($\mnew \sim 10^{16}$ GeV $<$ $\mpl \sim
10^{19}$ GeV), which could help alleviate hierarchy problems.
Nonetheless, the signatures of virtual black hole processes might be
observable in proton decay experiments, which may eventually provide 
a powerful experimental probe of quantum gravity and string theories. 

\bigskip 
\bigskip 
\centerline{\bf Acknowledgements} 
\medskip 

We greatly appreciate conversations with M. Duff, L. Everett,
S. Hawking, J. Liu, J. Lykken, and H. Reall; we particularly thank
D. Chung for important discussions. We are grateful to N. Arkani-Hamed
and G. Dvali for critical comments on an early version of the
manuscript.  Finally, we thank the Aspen Center for Physics, where
some of this work was carried out. This research was supported by a
Department of Energy grant and by the Physics Department at the
University of Michigan.

\bigskip 
\bigskip 
\centerline{\bf APPENDIX:} 
\centerline{\bf VIRTUAL BLACK HOLES AND SPACETIME FOAM} 
\bigskip 

In this Appendix, we present a version of the standard argument for
virtual black holes filling the vacuum and thereby producing a
spacetime foam. In this case, however, we generalize the calculation
for higher dimensions. We consider gravity to propagate in $4+n$
dimensions, so that $n$ is the number of large extra dimensions.
Notice that $n$ depends on the scale. On sufficiently large spatial
scales $n\to0$ and we must recover old (4-dimensional) Einstein
gravity; in this context we are interested in small spatial scales
where black holes must be $(4+n)$-dimensional. Gravity is controlled
by the action [22,23,26] 
\be 
I[g] = - {1 \over 16 \pi G_n} \int_M R \, (-g)^{1/2} d^{4+n}x - 
{1 \over 8 \pi G_n} \int_{\partial M} K \, (-h)^{1/2} d^{3+n}x \,  
+ C[h] \, , 
\ee 
where $G_n$ is the gravitational constant in $(4+n)$ dimensions and
$R$ is the Ricci scalar for the metric $g_{ab}$, which is defined on
the spacetime $M$.  The spacetime boundary $\partial M$ has the
induced metric $h_{ab}$. The quantity $K$ is the trace of the second
fundamental form on the boundary $\partial M$ and $C[h]$ is a
functional of $h$ defined so that the action of Minkowski space
vanishes.  Extremization of this action for fixed metric on the
boundary leads to the Einstein equations for $g_{ab}$ in $M$.

The path integral for gravity is
\be
Z \sim \int {\cal D}[g] \, {\rm e}^{i I[g]} \, , 
\ee
where the integral is taken over all metrics $g$. Our goal is to
investigate how black holes contribute to any amplitude in quantum
gravity. We first assume that this integral can be approximated by 
the usual Euclidean continuation.  The action for a single 
Schwarzschild black hole of mass $m$ is then given by
\be 
I_1 \sim {m^{2+n} \over \mpln^{2+n} } \, , 
\ee
where $\mpln$ is the Planck mass in $(4+n)$ dimensions ($\mpln 
\approx \mnew$). This equation should also contain additional 
geometrical factors, but these are of order unity and convention
dependent (depending on how the mass $m$ is defined). Ignoring
interactions between the black holes, we find the action for a
collection of $N$ black holes to be
\be 
I_N \sim {N m^{2+n} \over \mpln^{2+n} }  \, .
\ee
In the path integral, the black holes are indistinguishable; 
each is independent of the others and can be positioned anywhere 
in space.  Since $N$ is undetermined, we can evaluate $Z$ in a 
box of volume $V_n$ (in $3+n$ spatial dimensions) to obtain the result 
\be 
Z \sim \int_0^\infty dm \sum_{N=0}^\infty 
\exp[-4 \pi N m^{2+n} / \mpln^{2+n}] \, {1 \over N!} \, 
\Bigl[ {V_n \over \lpln^{3+n}} \Bigr]^N \, .
\ee
The factor of $V_n$ comes from accounting for the black holes 
being anywhere in the box, and the factor of $1/N!$ arises 
from their indistinguishability. 

The combination of these results thus defines a probability 
distribution for having $N$ black holes with mass $m$. Elementary 
calculations yield the corresponding expectation values for the 
number density $n_{\rm bh}$ of black holes and for the black hole 
mass $m_{\rm bh}$, i.e., 
\be 
\langle n_{\rm bh} \rangle \sim {V_n \over \lpln^{3+n}} 
\qquad {\rm and} \qquad \langle m_{\rm bh} \rangle \sim \mpln \, , 
\ee
where $\lpln$ is the Planck length in the $(4+n)$-dimensional
spacetime. As before in the case of 4 dimensions, we find that
spacetime must be filled with tiny Planck mass black holes with a
density of roughly one per Planck volume. These microscopic virtual
black holes live roughly for one Planck time. In this generalized
case, however, the black holes are $(4+n)$-dimensional objects and the
Planck mass, the Planck volume, and the Planck time are now given by
the new (lower) energy scale $\mpln$ (where $\mpln \approx \mnew$).
This picture of the spacetime vacuum is thus a generalization of the
spacetime foam for the case of $(4+n)$ dimensions.

\bigskip 
\newpage 
\centerline{\bf REFERENCES} 
\medskip 

\medskip\par\pp{[1]} 
I. Antoniadis, {\sl Phys. Lett.} {\bf B246}, 377 (1990).

\medskip\par\pp{[2]} 
J. Lykken, {\sl Phys. Rev.} D {\bf 54}, 3693, hep-th/9603133 (1996). 

\medskip\par\pp{[3]} 
K. R. Dienes, E. Dudas, and T. Gherghetta, {\sl Phys. Lett.} {\bf B436}, 
55 (1998); G. Shiu and S. H. Tye, {\sl Phys. Rev.} D {\bf 58}, 106007 (1998); 
P. C. Argyres, S. Dimopoulos, and J. March-Russell, {\sl Phys. Lett.} 
{\bf B441}, 96 (1998). 

\medskip\par\pp{[4]} 
J. Ellis, A. E. Faraggi, and D. V. Nanopoulos, 
{\sl Phys. Lett.} {\bf B419}, 123 (1998). 

\medskip\par\pp{[5]} 
N. Arkani-Hamed, S. Dimopoulos, and G. Dvali, {\sl Phys. Lett.} 
{\bf B429}, 263 (1998); {\sl Phys. Rev.} D {\bf 59}, 0806004 (1999);
I. Antoniadis, N. Arkani-Hamed, S. Dimopoulos, and G. Dvali, 
{\sl Phys. Lett.} {\bf B436}, 257 (1999). 

\medskip\par\pp{[6]} 
Z. Kakushadze, {\sl Nucl. Phys.} {\bf B552}, 3 (1999). 

\medskip\par\pp{[7]} 
T. Banks, M. Dine, and A. E. Nelson, hep-th/9903019 (1999). 

\medskip\par\pp{[8]} 
T. Banks and W. Fischler, hep-th/9906038 (1999). 

\medskip\par\pp{[9]} 
R. Casadio and B. Harms, hep-th/0004004 (2000). 

\medskip\par\pp{[10]} 
R. Emparan, G. T. Horowitz, and R. C. Myers, hep-th/0003118 (2000). 

\medskip\par\pp{[11]} 
N. Arkani-Hamed and M. Schmaltz, hep-ph/9903417 (1999);
N. Arkani-Hamed, Y. Grossman, and M. Schmaltz, hep-ph/9909411 (1999).  

\medskip\par\pp{[12]} 
P. Langacker, {\sl Phys. Rep.} {\bf 72}, 186 (1984). 

\medskip\par\pp{[13]} 
Super-Kamiokande Collaboration, 
{\sl Phys. Rev. Lett.} {\bf 81}, 3319 (1998); 
{\sl Phys. Rev. Lett.} {\bf 83}, 1529 (1999).

\medskip\par\pp{[14]} 
Ya. B. Zeldovich, Phys. Lett. {\bf A59}, 254 (1976);  
Ya. B. Zeldovich, Sov. Phys. JETP, {\bf 45}, 9 (1977). 

\medskip\par\pp{[15]} 
S. W. Hawking, D. N. Page, and C. N. Pope, Phys. Lett. {\bf B86}, 
175 (1979). 

\medskip\par\pp{[16]} 
D. N. Page, Phys. Lett. {\bf B95}, 244 (1980). 

\medskip\par\pp{[17]} 
J. Ellis, J. S. Hagelin, D. V. Nanopoulos, and K. Tamvakis, 
Phys. Lett. {\bf B124}, 484 (1983). 

\medskip\par\pp{[18]} 
F. C. Adams, G. Laughlin, M. Mbonye, and M. J. Perry, 
{\sl Phys. Rev.} D {\bf 58}, 083003 (1998). 

\medskip\par\pp{[19]} 
N. Arkani-Hamed, S. Dimopoulos, and G. Dvali, 
{\sl Scientific American}, August (2000). 

\medskip\par\pp{[20]} 
L. Randall and R. Sundrum, {\sl Phys. Rev. Lett.} {\bf 83}, 4690 (1999). 

\medskip\par\pp{[21]} 
J. Lykken and L. Randall, JHEP 0006, 014, hep-th/9908076 (2000). 

\medskip\par\pp{[22]} 
S. W. Hawking, Nucl. Phys. {\bf B144}, 349 (1978);
G. W. Gibbons and S. W. Hawking, Phys. Rev. D {\bf 15}, 2752 (1977). 


\medskip\par\pp{[23]} 
M. J. Perry, in {\sl Unification of Elementary Forces and Gauge Theories}, 
ed. D. B. Cline and F. E. Mills, p. 485 (Harwood, London, 1977). 

\medskip\par\pp{[24]}
A. Sen, Mod. Phys. Lett. {\bf A10}, 2081 (1995). 

\medskip\par\pp{[25]} 
A. P{\'e}rez-Lorenzana, hep-ph/0008333 (2000). 

\medskip\par\pp{[26]} 
R. C. Myers and M. J. Perry, Ann. Phys. {\bf 172}, 304 (1986); 
F. R. Tangherlini, Nuovo Cimento {\bf 27}, 636 (1963). 

\medskip\par\pp{[27]} 
G. Dvali, private communication (2000). 

\medskip\par\pp{[28]} 
F. C. Adams and G. Laughlin, Rev. Mod. Phys. {\bf 69}, 337 (1997). 

\end{document}